\documentclass[twocolumn,aps,prl,groupedaddress]{revtex4}
\usepackage{amssymb}
\usepackage{amsmath}
\usepackage{color}
\usepackage{graphicx}
\usepackage{marvosym}

\newcommand{\be}{\begin{equation}}
\newcommand{\ee}{\end{equation}}
\newcommand{\nn}{\nonumber\\}
\newcommand{\HH}{{\cal H}}

\newcommand{\p}{\partial}
\newcommand{\la}{\langle}
\newcommand{\ra}{\rangle}
\newcommand{\lb}{\left[}
\newcommand{\rb}{\right]}
\newcommand{\lp}{\left(}
\newcommand{\rp}{\right)}

\newcommand{\sgn}{{\rm sgn}\,}

\renewcommand{\Re}{{\rm Re}\,}
\renewcommand{\vec}[1]{{\bf #1}}

\begin{document}

\title{Large optical conductivity of Dirac semimetal Fermi arc surfaces states}
\author{Li-kun Shi$^{1}$ and Justin C. W. Song$^{1,2}$}
\affiliation{$^1$Institute of High Performance Computing, Agency for Science, Technology, \& Research, Singapore 138632}
\affiliation{$^2$Division of Physics and Applied Physics, Nanyang Technological University, Singapore 637371}

\begin{abstract}
Fermi arc surface states, a hallmark of topological Dirac semimetals, can host carriers that exhibit unusual dynamics distinct from that of their parent bulk. Here we find that Fermi arc 
carriers in intrinsic Dirac semimetals possess a strong and anisotropic light matter interaction. 
This is characterized by a large Fermi arc optical conductivity when light is polarized transverse to the Fermi arc; when light is polarized along the Fermi arc, Fermi arc optical conductivity is significantly muted. The large surface spectral weight is locked to the wide separation between Dirac nodes and persists as a large Drude weight of Fermi arc carriers when the system is doped.  
As a result, large and anisotropic Fermi arc conductivity provides a novel means of optically interrogating the topological surfaces states of Dirac semimetals. 
\end{abstract}

\pacs{pacs}
\maketitle

\begin{figure}[t!] \includegraphics[width=\columnwidth]{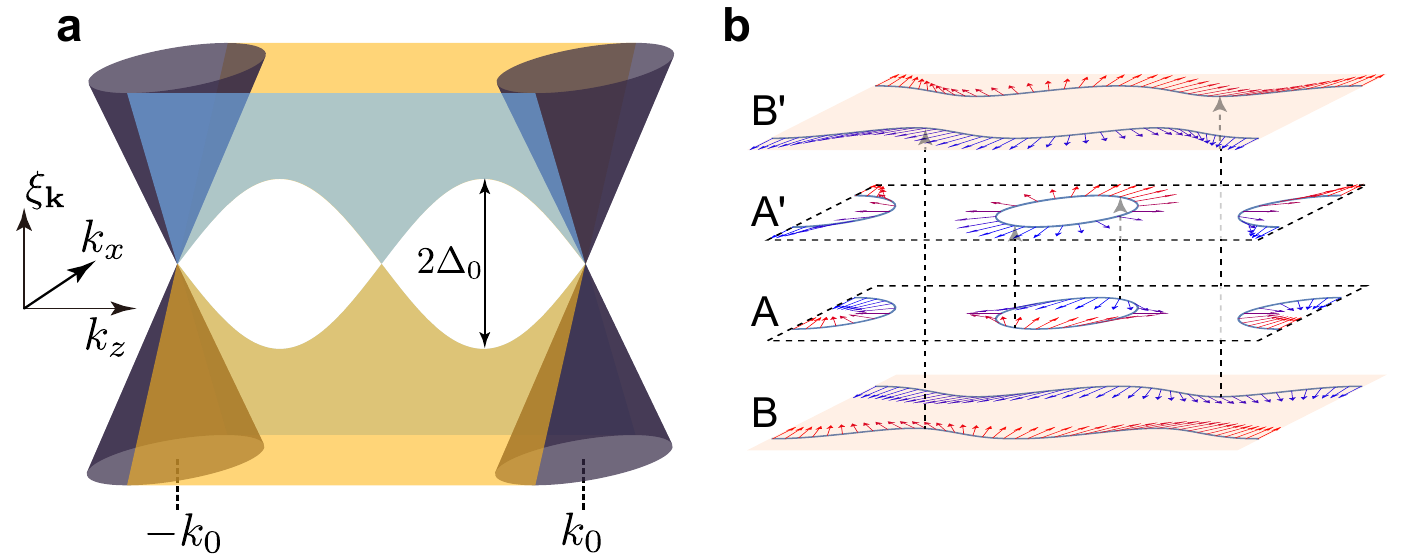}
\caption{(a) Schematic dispersion for surface states (blue and yellow) between the Dirac cones (purple) in three-dimensional topological Dirac semimetals; surface is terminated normal to $\hat{\vec y}$. Upper and lower branches of surface states are degenerate at $(k_x,k_z) =0$ and at the Dirac points $(k_x,k_z) = (0, \pm k_0)$, while gapped elsewhere with gap characterized by $\Delta_{\vec k}$, see text. (b) Particle-hole excitation between surface states at different energies. Transitions from $A \rightarrow A'$ ($B \rightarrow B'$) correspond to photon energy $\hbar \omega < 2 \Delta_0 $ ($\hbar \omega > 2 \Delta_0 $). Red and blue arrows display the spin textures. }
\label{fig1}
\end{figure}

Three-dimensional topological Dirac semimetals (DSMs) possess four-fold degenerate band touchings in the bulk (Dirac points) that are stabilized by the underlying crystal symmetries~\cite{wang,yang,neupanecd}. 
Of special interest are the unusual surface states in DSMs, which are localized on particular exposed crystal surfaces even in the presence of bulk metallic states. These Fermi arc surface states feature several interesting properties including spin-momentum locking~\cite{neupanecd,xuarc}, and possess a dispersion that spans the large expanse of momenta between the Dirac nodes (Fig.~\ref{fig1}a)~\cite{neupanecd,jeoncd,liu,liangcd,neupaneche,xuna,sergey,akrap,xuarc,burkov}. As a result, carriers on the Fermi arc surface states exhibit markedly different dynamics from that of carriers in the bulk --- a hallmark of the unconventional Fermiology of DSMs~\cite{potter,moll}.

\begin{figure}[t!]
\includegraphics[width=\columnwidth]{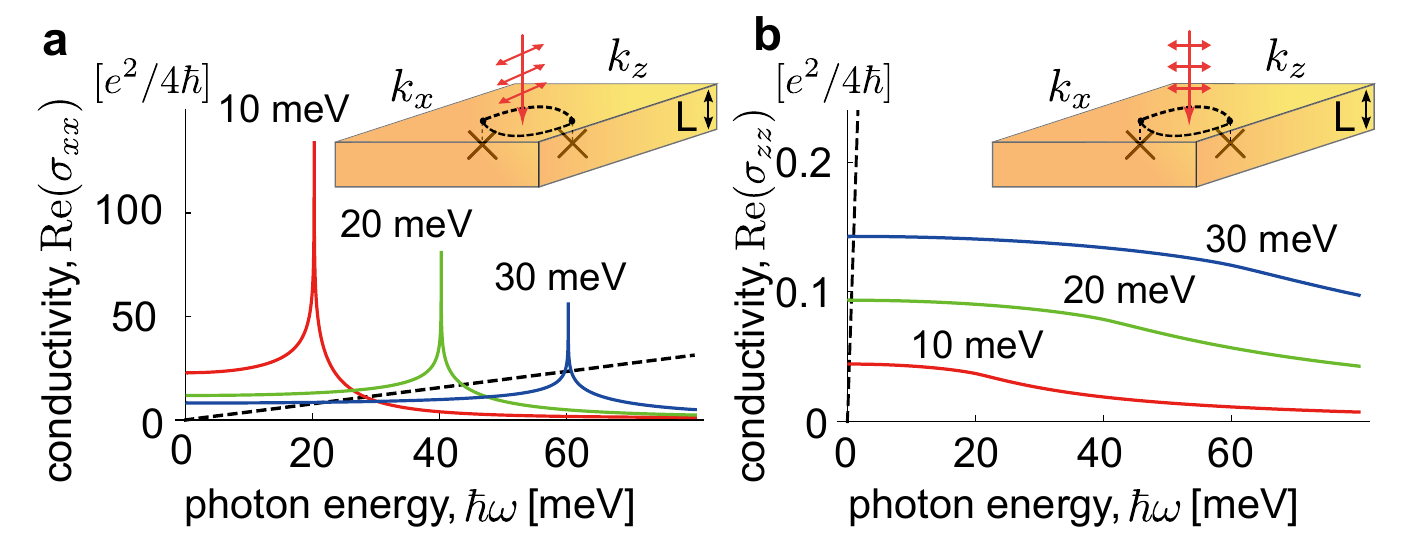}
\caption{
Real part of interband conductivity ${\rm Re}(\sigma_{xx})$ [solid lines in (a)] and ${\rm Re} (\sigma_{zz})$ [solid lines in (b)] from inter Fermi arc state transitions obtained from 
Eq.~(\ref{eq:interbanddimensionless}) at $\mu =0$. 
For comparison, bulk inter-band transitions for a $500~{\rm nm}$ thin slab, with bulk carrier velocity $A_x = v$ and $A_z = 250~{\rm meV \cdot nm}$ ($\approx 3.8 \times 10^5~{\rm m \cdot s^{-1} }$), are shown in dashed lines (see~\cite{carbotte,supp} for detailed calculations). We have chosen units $e^2/4\hbar$ for the optical conductivity of graphene as a reference. Parameters used: $v = 450~{\rm meV \cdot nm}$ ($\approx 6.9 \times 10^5~{\rm m \cdot s^{-1} }$) and $k_0 = 1.5~{\rm nm^{-1}}$. Red, green, and blue correspond to $\Delta_0 = 10\, , 20\, , 30\, {\rm meV}$ respectively.}
\label{fig2}
\end{figure}

Here we argue that Fermi arc surface states in DSMs can mediate a large optical response. In particular, for an undoped DSM with Dirac nodes at $k_z = \pm k_0$ (Fig.~\ref{fig1}), we find that the interband optical conductivity associated with optical absorption in Fermi arc surface states 
can attain large values of several tens of $e^2/h$ when incident light is polarized along the $\hat{\vec{x}}$ direction 
(Fig.~\ref{fig2}a). These values are far larger than those of other gapless two-dimensional electron gases (e.g., undoped graphene~\cite{nair}).  
Further,  interband optical conductivity displays a distinctive peak like frequency dependence which does not vanish at small frequencies (Fig.~\ref{fig2}a) arising from the double-arch-shaped dispersion of the DSM Fermi arc states (Fig.~\ref{fig1}). This contrasts with the vanishing optical conductivity expected from the DSM bulk~\cite{carbotte}.

Large DSM surface absorption along $\hat{\vec{x}}$ stems from the wide range of momenta between $-k_0 <k_z <k_0$ that DSM Fermi arc surface states span. As a result, a large set of particle-hole transitions occur, leading to large surface absorption values. Additionally, since DSM Fermi arc surface states arise from a band inversion between bands of different parity, they possess a canted spin texture. As we explain below, this spin texture mediates contrasting optical selection rules for interband electron-hole transitions when incident light is polarized in $\hat{\vec{x}}$ or $\hat{\vec{z}}$ directions. As a result, when incident light is polarized along $\hat{\vec{z}}$, we find absorption resulting from transitions between Fermi arc surface states can be 100-1000 times smaller, displaying a large anisotropy of the Fermi arc surface states (Fig.~\ref{fig2}b). 

When the DSM is doped, the large values of DSM surface absorption persists in the form of a Drude weight along $\hat{\vec{x}}$; consequently this mediates a large intraband (Drude) absorption. The total optical weight for both intra- and inter-band absorption along $x$ is conserved and we obtain an explicit expression for the spectral weight that is directly dependent on the separation between the Dirac nodes, $2k_0$. Since the large DSM surface absorption is locked transverse to the vector separating the pair of Dirac nodes, DSM surface absorption can be used as a probe of the crystal orientation. Its distinctive frequency dependence departs from that expected from bulk states and provides an optical means of probing the dynamics of the DSM Fermi arc surface state carriers.

\vspace{2mm}
{\bf Effective hamiltonian and surface states---} 
We begin by considering three-dimensional topological Dirac semimetals (DSMs)~\cite{wang,yang,burkov} which possess rotational symmetry along the $k_z$ axis, and a pair of Dirac points at $\pm \vec q_0 =  (0,0,\pm k_0)$
within the Brillouin zone~\cite{neupanecd,jeoncd,liangcd,liu,neupaneche,xuna,sergey,akrap,xuarc}. These can be described by
a minimal $4 \times 4 $ Hamiltonian~\cite{wang,yang,burkov}, $\mathcal{H} = \HH_0 + \HH' $ with
\be
\HH_0 = m_{\vec q} \vec s_0 \boldsymbol{\tau}_3 + v (k_x \vec s_3 \boldsymbol{\tau}_1 - k_y \vec s_0 \boldsymbol{\tau}_2),
\quad
\mathcal{H}'\sim  \vec s_1,\vec s_2, 
\label{eq:generalhamiltonian}
\ee
where $\vec q = (k_x,k_y,k_z)$ is taken about the $\Gamma$ point, $\vec{s}_0$ ($\boldsymbol{\tau}_0$) is an identity matrix, and $\vec s_{1,2,3}$ ($\boldsymbol{\tau}_{1,2,3}$) are Pauli matrices representing the spin (orbital) degrees of freedom. Here, mass takes the form $m_{\vec q} = - m_0 + m_1  k_z^2 $ along the $k_z$ axis, with $m_0 >0 $ capturing band inversion~\cite{wang,yang} between the Dirac points $\pm k_0 = \pm \sqrt{m_0/m_1}$. 

Since $\HH_0$ in Eq.~(\ref{eq:generalhamiltonian}) possesses only diagonal $\vec{s}_0,\vec s_3$ terms, $\HH_0$ is block diagonal in spin space; the pair of $2\times2$ blocks (denoted $s= \pm 1$) are related via time-reversal symmetry (TRS)~\cite{wang,yang,xuarc}.
In contrast, $\HH' \sim \vec s_1,\vec s_2$ is purely off-diagonal, coupling the $s=\pm 1$ blocks. Further, it is weak with $\mathcal{H}' \sim \mathcal{O}(k^3)$, and vanishes along the  $k_z$ axis due to rotational symmetry (RS)~\cite{wang,yang,neupanecd}. 
As a result, a pair of degenerate Dirac nodes emerge at $\pm \vec q_0$ where scattering between $s=+1$ and $s=-1$ blocks vanish. $\mathcal{H}$ captures the essential low-energy physics of DSMs, and has been recently used to describe several DSM systems including Na$_3$Bi and Cd$_3$As$_2$ \cite{jeoncd,liu}.

Of particular interest are surface states that develop on certain exposed surfaces due to the band-inversion of conduction and valence bands with different parities between the Dirac nodes. To illustrate these states, we will consider a DSM in the region $y\geq 0$ described by Eq.~(\ref{eq:generalhamiltonian}); with vacuum occupying $y<0$ . For each fixed $\vec k = (k_x,k_z)$ between the Dirac points projected onto the $x$-$z$ plane, each block $s=\pm 1$ of Eq.~(\ref{eq:generalhamiltonian}) describes a one-dimensional Dirac system with a sign changing mass as a function of $y$~\cite{burkov,yang}. As a result, Jackiw-Rebbi type surface states emerge, and are localized about $y=0$:
\be
 | \Psi_{ \vec k, s} (y) \ra = C e^{ - y/\lambda_{\vec k} } u_{s},
 \label{eq:surfacestates}
 \ee
where $u_{s=+1} = (1,1,0,0)^T$,  $u_{s=-1} = (0,0,1,1)^T$, decay length $\lambda_{\vec k} =  - v/ m_{\vec k}>0$~\cite{supp}, and $C$ is a normalization constant. Here $u_s$ are $\vec k$-independent block-spinors arising from the $s = \pm 1$ block in Eq.~(\ref{eq:generalhamiltonian}) respectively~\cite{supp}. The surface states in Eq.~(\ref{eq:surfacestates}) only exist between $k_z = \pm k_0$ due to wavefunction normalizability~\cite{supp}. Indeed, when $k_z \to  \pm k_0$, $\lambda_{\vec k}$ diverges, indicating that $\Psi_{ \vec k, s} (y)$ merges into the bulk.

Using the surface state wavefunctions in Eq.~(\ref{eq:surfacestates}) as a basis $\{| \Psi_{ \vec k, s=+1}(y) \ra, | \Psi_{ \vec k, s=-1} (y)\ra \}$, we write a Hamiltonian describing the surface electronic behavior as
\be
\HH_{\rm surf} =  v k_x \boldsymbol{\sigma}_z + \Delta_{\vec k} \boldsymbol{\sigma}_x ,
\label{eq:surfacehamiltonian}
\ee
where the first term is obtained directly from Eq.~(\ref{eq:generalhamiltonian}):
 $ v k_x \boldsymbol{\sigma}_z = \la \Psi_{ \vec k, s}(y) | vk_x \vec s_3 \boldsymbol{\tau}_1| \Psi_{ \vec k, s}(y) \ra$ describes the linear dispersion of the surface states. 
The second term $\Delta_{\vec k} \boldsymbol{\sigma}_x$ describes inter-block mixing [e.g., intrinsic $ \Delta_{\vec k} \boldsymbol{\sigma}_x = \la \Psi_{ \vec k, s}(y) |\mathcal{H}'| \Psi_{ \vec k, s'}(y) \ra$ for $s\neq s'$]. We note that scattering induced by defects or a surface potential that break the crystalline order may also lead to a finite $\Delta_{\vec k}$~\cite{potter}. As a result, below we will use a phenomenological model for $\Delta_{\vec k}$.

Crucially, the spinors $\{| \Psi_{ \vec k, s=+1}(y) \ra, | \Psi_{ \vec k, s=-1} (y)\ra \}$
decouple at zero nodes of $\Delta_{\vec k}$ as governed by the symmetries of the system. For example, at $\vec k=0$, $\Delta_{\vec k=0}$ vanishes since $| \Psi_{ \vec k=0, s}(y) \ra$ and $|\Psi_{ \vec k=0, -s} (y)\ra$ are a Kramers pair related by TRS
. To see this explicitly, note $\Delta_{\vec k=0} = \la \Psi_{\vec k=0, s} | \HH(\Gamma)| \Psi_{\vec k=0, -s} \ra = \la \Theta^2  \Psi_{\vec k=0, s} | \HH(\Gamma)| \Theta \Psi_{\vec k=0, s} \ra  = - \Delta_{\vec k=0} = 0$~\cite{supp}. 
Similarly, in the presence of RS about the $k_z$ axis~\cite{wang,yang} on the surface, $\Delta_{\vec k}$ also vanishes at the two projected Dirac points $\pm \vec k_0 = (0, \pm k_0)$. Interestingly, while the zero node at ${\vec k}=0$ is robust against disorder/defects that preserve TRS, the nodes at $\pm \vec k_0$ are more fragile and can be gapped out by non-rotationally symmetric defects on the surface.

The eigenfunctions of Eq.~(\ref{eq:surfacehamiltonian}) can be readily obtained as $\psi_{\vec k}^+ = [ \sin (\phi_{\vec k}/2), \cos (\phi_{\vec k}/2) ]$ and $ \psi_{\vec k}^- = [ \cos (\phi_{\vec k}/2), - \sin (\phi_{\vec k}/2) ] $, with energy eigenvalue $\xi_{\vec k}^\pm = \pm \eta_{\vec k}$, where $\tan \phi_{\vec k} = \Delta_{\vec k}/v k_x $ \cite{gauge} and $\eta_{\vec k} =(\Delta_{\vec k}^2 + v^2 k_x^2 )^{1/2}$. Each of the components of $\psi_{\vec k}$ denote the wavefunction weight on the $s=\pm 1$ blocks respectively represented by $| \Psi_{ \vec k, +1} \ra$  and $| \Psi_{ \vec k, -1} \ra$. These surface states can be probed directly via angle-resolved photoemission spectroscopy (ARPES)~\cite{neupanecd,liu,neupaneche,xuna,xuarc,sergey}. Additionally, carriers moving along DSM states may also exhibit peculiar cyclotron orbits~\cite{potter} and can be probed via quantum oscillations~\cite{moll}.

\vspace{2mm}
{\bf Inter Fermi arc transitions ---}
As we now show, DSM surface states in Eq.~(\ref{eq:surfacehamiltonian}) exhibit a strong light-matter interaction. In the presence of normally incident light $\boldsymbol {\cal E} = \vec E \cos \omega t$ to the exposed surface, electron-hole transitions occur between the occupied and unoccupied surface states (Fig.~\ref{fig1}a,b). These can be described via the standard golden rule technique as
\be
W = \frac{2\pi}{\hbar} \sum_{\vec k} |M_{\vec k}|^2 \delta(\xi_{\vec k}^+ - \xi_{\vec k}^- - \hbar \omega) f (\xi_{\vec k}^-) [1 - f(\xi_{\vec k}^+)],
\label{eq:transitionrategeneral}
\ee
where $f (x) = [ e^{(x - \mu)/ k_B T} +  1 ]^{-1}$ is the Fermi function, with $\mu$ the chemical potential.

For clarity, we adopt a simplified model for $\Delta_{\vec k} = \Delta_0 \sin (\pi k_z/k_0)$ in Eq.~(\ref{eq:surfacehamiltonian}). This choice of $\Delta_{\vec k}$ captures the zero nodes at ${\vec k} =0$, as well as the two projected Dirac nodes at $(0, \pm k_0)$~\cite{supp}.
We note that other models for $\Delta_{\vec k}$ can also be used, however, we do not expect them to alter the qualitative conclusions discussed below.
Writing $\vec k \rightarrow \vec k - e \boldsymbol {\cal A}/ \hbar c$ in Eq. (\ref{eq:surfacehamiltonian}), with the vector potential satisfying $\boldsymbol{\mathcal{E}} = (1/c) \partial_t \boldsymbol{\mathcal{A}}$ 
and expanding to linear order in $\vec E$ yields the matrix elements
\be
M_{\vec k} = \frac{i e}{2 \hbar \omega} \la \psi_{\vec k}^+ | v_x E_x \boldsymbol{\sigma}_z  +   v_z \cos (\pi k_z / k_0) E_z \boldsymbol{\sigma}_x  | \psi_{\vec k}^- \ra,
\label{eq:matrixelement}
\ee
where $v_x = v$, and $v_z = \pi \Delta_0/ k_0 $.

Particle-hole excitations occur on the constant energy contours defined by $\eta_{\vec k} = \hbar \omega / 2$ in $k$-space since energy conservation demands $\delta(\xi_{\vec k}^+ - \xi_{\vec k}^- - \hbar \omega) = \delta(2 \eta_{\vec k} - \hbar \omega)$. These contours can be classified into two types: (A) $\hbar\omega < 2\Delta_0$ wherein particle-hole excitations occur along contours with multiple disconnected segments (see Fig.~1b), and (B) $\hbar \omega > 2\Delta_0$ where particle-hole excitations occur along two long singly-connected arcs (the so-called ``Fermi arcs'', see Fig.~1b). When $\hbar\omega = 2 \Delta_0$, when the particle-hole excitations transform from A- to B-type contours.

In order to keep track of the distinct topologies of A- and B-type contours, it will be convenient to define the particle-hole excitation spectrum as
\be
\rho(\omega,\nu) = \int_0^\infty R(\eta,\nu) \delta (2 \eta - \hbar \omega) d\eta, \quad R = \left|\frac{\partial(k_x,k_z)}{\partial(\eta,\nu)}\right|, \label{eq:particle-hole-spectrum}
\ee
with $\nu (\vec k)= \phi_\vec{k}$ for $\hbar \omega < 2\Delta_0$ describing A-type contours, and $\nu(\vec k) =\pi k_z/k_0$ for $\hbar\omega\geq 2\Delta_0$ denoting B-type contours; in both cases $-\pi<\nu<\pi$. This choice of $\nu (\vec k)$ results in simple forms for the Jacobian: $ R(\eta,\nu) = 2 D (\eta / \Delta_0) [ 1 - (\eta / \Delta_0)^2 \sin^2 \nu ]^{-1/2}$ for A-type contours ($\hbar \omega < 2\Delta_0$), and $ R(\eta,\nu) = 2 D [ 1 - ( \Delta_0 / \eta)^2 \sin^2 \nu ]^{-1/2}$ for B-type contours ($\hbar \omega \geq 2\Delta_0$), where $D = (2\pi)^{-2} (\Delta_0 /v_x v_z ) $~\cite{supp}. 

Using Eq.~(\ref{eq:particle-hole-spectrum}), we can write the transition rate, Eq.~(\ref{eq:transitionrategeneral}), in a compact form
\be
W (\omega)= \frac{2\pi}{\hbar} \int_{-\pi}^{\pi} d\nu|{\cal M}(\omega,\nu)|^2 \rho (\omega, \nu)  f (\xi^{-}) \big[1 - f(\xi^{+})\big],
\label{eq:surfacetransitionrate}
\ee
where the matrix elements ${\cal M} = {\cal M}_x + {\cal M}_z$ are
\be
{\cal M}_{x} = \frac{eE_x}{2 \hbar \omega} \frac{ v_x  \sin \nu}{\bar{\omega}^{\frac{1}{2}(1+\zeta)}} , \quad {\cal M}_{z}= \frac{eE_z}{2 \hbar \omega} \frac{ v_z   \cos \nu}{(1 - \bar{\omega}^{-2\zeta} \sin^2 \nu)^{-\frac{1}{2}}} ,
\label{eq:matrixelements}
\ee
where $\bar{\omega} = \hbar \omega / 2 \Delta_0$, and we have noted $\la \psi_{\vec k}^+ | \boldsymbol{\sigma}_x | \psi_{\vec k}^- \ra = \cos \phi_{\vec k}$ and $\la \psi_{\vec k}^+ | \boldsymbol{\sigma}_z | \psi_{\vec k}^- \ra =  \sin \phi_{\vec k}$. Here $\zeta(\bar{\omega})= \sgn (\bar{\omega} -1)$ tracks the A-type (where $\zeta = -1$) and B-type (where $\zeta =1$) contours that determine the particle-hole excitations.
Performing the integral in Eq.~(\ref{eq:particle-hole-spectrum}), we obtain 
\be
\rho(\omega, \nu) = \frac{1}{2\pi^2} \frac{\Delta_0}{v_x v_z} \frac{ \bar{\omega}^{\frac{1}{2} (1-\zeta) } }{ (1 - \bar{\omega}^{-2\zeta} \sin^2 \nu)^{\frac{1}{2}} }.
\ee
As we will see below, the distinct angular features of both $\cal M$ and $\rho$ lead to a large anisotropy in the inter-band absorption spectrum when the incident light field is polarized along $\hat{\vec{x}}$ vs $\hat{\vec{z}}$. 

Using Ohms' law, $\hbar \omega W(\omega) = \Re [ \sigma(\omega)] |\vec E|^2/2$, and Eq.~(\ref{eq:surfacetransitionrate}) we obtain the inter-band conductivity for $\boldsymbol {\cal E}$ polarized along $\hat{\vec{x}}$ and $\hat{\vec{z}}$ directions as
\be
{\rm Re} \big[ \sigma_{ii}(\omega) \big]= (e^2/\hbar)  \chi_{ii}(\omega) \Theta(\hbar \omega/2 - |\mu|), \quad (i=x,z),
\label{eq:interbandconductivity}
\ee
where we have approximated $f (\xi^{-}) \big[1 - f(\xi^{+})\big] = f ( - \hbar \omega / 2) \big[1 - f(\hbar \omega /2)\big] = \Theta ( \hbar \omega/2 -  |\mu| )$ at low temperatures $k_B T \ll \hbar \omega/2 $.  
Here $\chi_{xx} (\omega)$ and $\chi_{zz}(\omega)$ (the dynamical conductivity at $\mu=0$) are
\be  
\chi_{xx}(\omega) =  \frac{v_x}{v_z}  I_x(\omega),  \quad  \chi_{zz}(\omega) = \frac{v_z}{v_x}   I_z(\omega) ,
\label{eq:interbanddimensionless}
\ee
where $I_x(\omega) = \frac{\bar{\omega}^{-\frac{3}{2} (1+\zeta)} }{4\pi} \int_{-\pi}^{ \pi } (1 - \bar{\omega}^{-2\zeta} \sin^2 \nu)^{-\frac{1}{2}}
\sin^2 \nu   d \nu $, and $I_z(\omega) = \frac{\bar{\omega}^{ - \frac{1}{2} (1+ \zeta)}}{4\pi}  
\int_{-\pi}^{ \pi } (1 - \bar{\omega}^{-2\zeta} \sin^2 \nu)^{\frac{1}{2}} 
\cos^2 \nu   d \nu $.

Plotting Eq.~(\ref{eq:interbanddimensionless}) and Eq.~(\ref{eq:interbandconductivity}) in Fig.~\ref{fig2}a we find a large Fermi arc interband optical conductivity ${\rm Re}(\sigma_{xx})$ when light is polarized along $\hat{\vec{x}}$; for parameters see caption.
Indeed, even at relatively low frequencies, ${\rm Re}(\sigma_{xx})$ can be up to 20 times of that found in monolayer graphene~\cite{nair} [Fig.~2a]. Strikingly, for frequencies approaching $\hbar \omega = 2\Delta_0$, surface ${\rm Re}(\sigma_{xx})$ diverges. This peak structure in the frequency dependence of ${\rm Re}(\sigma_{xx})$ 
arises from a van Hove singularity when the particle-hole spectrum  
transitions between A- and B- type contours (Fig. 1a,b). Its frequency dependence contrasts with the linear frequency dependence of DSM bulk interband optical conductivity that which vanishes at low frequency~\cite{carbotte}.

At low frequencies, the large values of ${\rm Re}(\sigma_{xx})$ can even dominate over absorption in the bulk for moderately thin-slab samples. For illustration, we note that the bulk optical conductivity in DSM slabs can be written as $\Re [G_{xx}^{\rm bulk}(\omega) ]= (e^2/\hbar)(1/12\pi) (\hbar \omega L_y / A_z) \Theta(\hbar \omega/2 - \mu)$~\cite{carbotte,supp}, and is linearly dependent on the thickness $L_y$, with $A_z$ the bulk carrier velocity~\cite{liu,liangcd,neupanecd}. Taking $L_y = 500~{\rm nm}$ as an example, we find that the surface conductivity in the $x$ direction (solid line) overwhelms that of the bulk (dashed line) at low frequencies [Fig.~\ref{fig2}a]. 

For light polarized in the $\hat{\vec{z}}$ direction, the Fermi arc interband optical conductivity ${\rm Re}(\sigma_{zz})$ (Fig.~\ref{fig2}b) exhibits a significantly muted magnitude and frequency dependence from that found for ${\rm Re}(\sigma_{xx})$. To understand this we note that the ratio between $\chi_{xx}/\chi_{zz}$ scales as the square of their velocities $(v_x/v_z)^2 = (v k_0 / \pi \Delta_0)^2$ [see Eq.~(\ref{eq:interbanddimensionless})], yielding a large ratio between $\chi_{xx}/\chi_{zz} \sim 100-1000$. This produces small ${\rm Re}(\sigma_{zz})$ values and a linear dichroism for transitions between Fermi arc states~\cite{dichroism}. In contrast to ${\rm Re}(\sigma_{xx})$ above, we find that bulk interband optical conductivity (dashed line) prevails over surface Fermi arc (solid lines) when light is polarized along $\hat{\vec z}$ [Fig.~\ref{fig2}b].

DSM surface state linear dichroism arises from the combined action of the canted spin texture of $\psi_{\vec k}$ on the surface, as well as the large spectrum of particle-hole excitations that are supported in the wide DSM surface state phase space, $-k_0< k_z < k_0$. Indeed, weak scattering between $s= \pm 1$ blocks encoded by a small $\Delta_0$, leads to a slow $v_z$ velocity concomitant with a giant dichroism. The linear dichroism of the DSM surface states (locked to the orientation of the Dirac nodes along $\hat{\vec z}$) provide a means of determining crystal axis configuration of the sample.

We note that the presence/absence of rotational symmetry on the surface can alter the nodal structure close to the projected Dirac points at $k_z =\pm k_0$; when rotational symmetry on the surface is broken, the nodes at $k_z =\pm k_0$ are gapped out. However, the node at $\vec k=0$ is protected by TRS. As a result, we expect the qualitative features of ${\rm Re}(\sigma_{xx})$ to persist even when rotational symmetry on the surface is broken.

\begin{figure}[t!]	\includegraphics[width=\columnwidth]{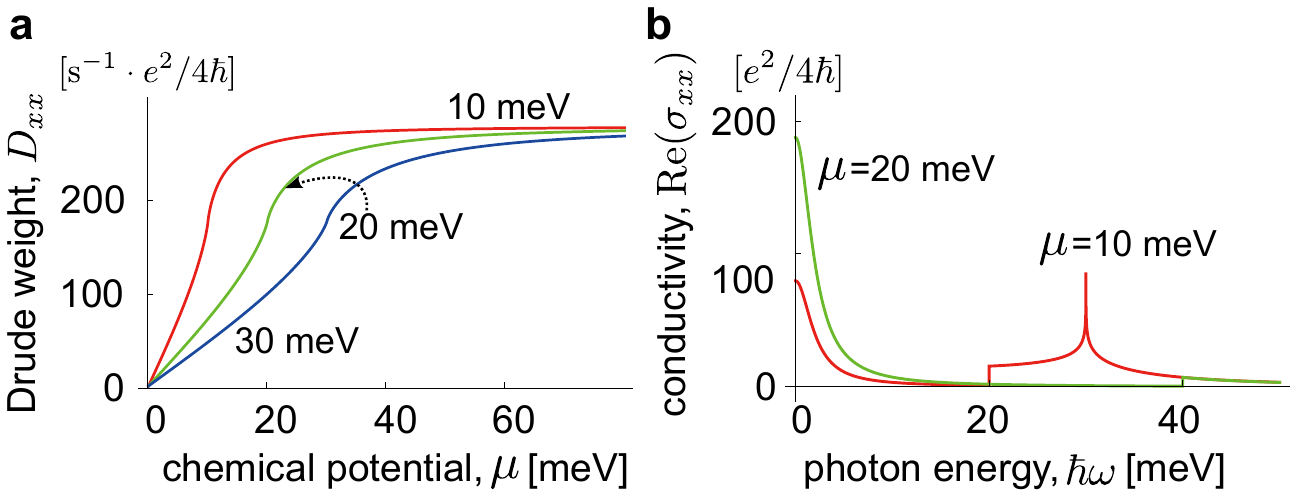}
\caption{ 
(a) Drude weight $D_{xx}$ obtained from Eq.~(\ref{eq:drudeweight}), with parameters used same as Fig.~\ref{fig2}; (b) Real part of conductivity $\sigma_{xx}$, including both inter- and intra- band parts, at different chemical potentials $\mu$, where we have used $\Delta_0 = 15~{\rm meV}$ and scattering time $\tau = 10~{\rm ps}$ for illustration. 
}
\label{fig3}
\end{figure}

\vspace{2mm}
{\bf Fermi arc intraband conductivity --- }At finite chemical potential, $\mu \neq 0$, low frequency optical transitions described in Eq.~(\ref{eq:interbandconductivity}) are Pauli blocked. As we now argue, the large conductivity along $\hat{\vec x}$ persists in the Drude conductivity. We capture this by a simple Drude model
\be
\sigma_{xx}^{\rm D} (\omega) =  \frac{ D_{xx}}{- i \omega + 1 / \tau}, \quad D_{xx} = -e^2 \sum_{\vec k}u_{\vec k x}^2 \frac{\p f(\xi_{\vec k})}{ \p \xi_{\vec k}},
\label{eq:intrabandconductivity}
\ee
where $\tau$ is the scattering time, and $D_{xx}$ is the Drude weight. Here $ u_{\vec k x} = \la \psi_{\vec k} | \partial \HH_{\rm surf} / \p (\hbar k_x)| \psi_{\vec k} \ra$ is the band velocity in the $\hat{\vec x}$ direction.

For the degenerate limit, $k_B T \ll \mu$ at low temperatures, we use $[ - \p f(\xi^\alpha_{\vec k}) / \p \xi_{\vec k} ] = \delta(\eta_{\vec k} - \mu)$. Following the same procedure above in Eq.~(\ref{eq:particle-hole-spectrum})-(\ref{eq:interbanddimensionless}), we obtain 
\be
D_{xx} = \frac{e^2}{\hbar} \frac{v_x}{ v_z } \frac{ 2 \Delta_0}{\pi \hbar}  {\cal I}_x (\mu),  \quad \bar{\mu} = \mu / \Delta_0
\label{eq:drudeweight}
\ee
where ${\cal I}_x(\mu) {} = \tfrac{\bar{\mu}^{(1-\zeta)/2}}{4\pi} \int_{-\pi}^{\pi} [1-\bar{\mu}^{-(1+\zeta)}\sin^2 \nu]/ [1-\bar{\mu}^{-2\zeta}\sin^2 \nu]^{1/2} d \nu$, and $\zeta(\bar{\omega})= \sgn (\bar{\omega} -1)$ as defined in Eq.~(\ref{eq:matrixelements}). Here we have used $(u_{\vec k x})^2 = (v_x/\hbar)^2 \cos^2 \phi_{\vec k}$.

In Fig.~\ref{fig3}a, we plot the Drude weight obtained from Eq.~(\ref{eq:drudeweight}) as a function of $\mu$. The Drude weight increases sharply from $\mu=0$, and saturates to a large value $D_{xx} \to D_0 =  (e/\pi \hbar)^2 v_x k_0 $ for $\bar{\mu}\gg 1$. Here we have noted that ${\cal I}_x(\mu \rightarrow +\infty) \rightarrow 1/2$. Interestingly, the large saturated value $D_0$ is independent of $\Delta_0$, and depends directly on the distance between the Dirac nodes, $k_0$. We note that the magnitude of $k_0$ is determined by the strength of band inversion $m_0$ between the opposite parity bands in the bulk as well as the $k_z$ band dispersion parameter $m_1$. Similarly, $D_0$ captures the optical weight that is redistributed to the surface (states) due to bulk band inversion~\cite{burkov,yang,wang}. Since the value of $k_0$ tracks the transition from normal insulator to a topological Dirac semimetal~\cite{yang}, the weight $D_0$ accumulated on the surface Fermi arcs can be used to optically probe this quantum phase transition.

Large values of $k_0$ are also responsible for the strong inter-band optical conductivity along $\hat{\vec x}$ described above, see Eq.~(\ref{eq:interbanddimensionless}). Indeed, directly summing the Drude and the inter-band conductivities along $\hat{\vec x}$ in Eq.~(\ref{eq:interbanddimensionless}) and Eq.~(\ref{eq:drudeweight}), we obtain an approximate partial sum rule
\be
\int_{0}^{\Omega} \Re [\sigma^D_{xx}(\omega)] d\omega + \int_{0}^{\Omega} \Re [\sigma_{xx}(\omega)] d \omega = \cal{S},
\label{eq:sum}
\ee
where we have taken $\Omega \gg 1/\tau, \hbar/\Delta_0, \hbar/\mu $, and 
$
{\cal S} = (e^2/\hbar)( v_x k_0 /2 \pi \hbar)$
is a constant that depends on $k_0$, the distance between the Dirac nodes. In obtaining the value of $\mathcal{S}$, we have noted that $ \int_{0}^{\Omega} \Re [\sigma_{xx}(\omega)] d \omega = c \int_{2\mu}^{\Omega} I_{x}(\mu') d \mu' = c \Delta_0 [{\cal I}_x(\Omega/2)- {\cal I}_x(\mu) ] $, where $c = (e^2/\hbar^2) (v_x/v_z)$. In the last line, we have used $\partial_{\mu} {\cal I}_x (\mu)= \Delta_0^{-1} I_x (2\mu)$~\cite{supp}. We also noted that the frequency integral of the intra-band conductivity yields $ \int_{0}^{\Omega} \Re [\sigma_{xx}^D (\omega)] d \omega = (\pi/2)D_{xx} =c\Delta_0 {\cal I}_x(\mu) $. 

To illustrate the conservation of the optical weight $\mathcal{S}$ in Eq.~(\ref{eq:sum}), we plot the inter- and intraband (Drude) conductivity obtained from Eq.~(\ref{eq:interbandconductivity}) and (\ref{eq:intrabandconductivity}) as a function of $\omega$ in Fig.~\ref{fig3}b. Note that as $\mu$ is tuned from $10 \to 20 \, {\rm meV}$, parts of the red curve associated with interband conductivity become Pauli blocked. As expected from Eq.~(\ref{eq:sum}), the missing area is transferred to the intraband conductivity, and is captured as an increased Drude weight, $D$ and a larger (green) peak height for the intraband (Drude) conductivity in Fig.~\ref{fig3}b. Eq.~(\ref{eq:sum}) is a powerful tool to analyze the low-energy spectroscopic features of the DSM surface states. Indeed, similar sum rule analyses have been used successfully to study two-dimensional surface states e.g., graphene~\cite{basovrmp} and topological insulator surface states~\cite{post2015}. 

Before closing, we briefly outline the conditions that favor the observation of large surface absorption in DSMs. For example, we note that the regime in which inter-Fermi arc transitions occur is limited by the bandwidth of the Fermi arc surface states $\sim v_x k_0$ which also directly determines the optical weight $\mathcal{S}$. Hence both wide $k_0$ as well as fast $v_x$ are favorable. In addition, we note that when DSM bands possess particle-hole (p-h) asymmetry (e.g. via $\epsilon_{\vec q}$ in Eq.~(\ref{eq:surfacehamiltonian})), the particular distribution of occupied and unoccupied surface states that participate in absorption are directly impacted. For example, p-h asymmetry can induce additional Pauli blocking that can suppress the contribution of inter-Fermi arc transitions to surface absorption and skew the linear dichroism discussed above. For $\epsilon_{\vec q} \ll \Delta_{\vec q}$, the effects of p-h asymmetry are negligible. Lastly, moderately thin samples can suppress the contribution for bulk transitions, allowing the surface absorption mediated by the Fermi arcs to be observed clearly.

In addition to the inter Fermi arc transitions (discussed above), and bulk-bulk interband transitions~\cite{carbotte,supp}, we note that inter Fermi arc-bulk transitions may also occur. These transitions will further add to the already large surface absorption from inter Fermi arc transitions described above. However, we expect these depend sensitively on the details of the dispersion on the surface. For example, when rotational symmetry on the surface is broken, the nodes at the projected Dirac points at $k_z = \pm k_0$ are gapped out. As a result, the inter Fermi arc-bulk transitions for intrinsic DSMs will exhibit a gap. In contrast, inter Fermi arc transitions remain gapless since the node at $\vec k =0$ is protected by TRS.

Fermi arc carriers can possess a strong light-matter interaction, characterized by a large optical conductivity, peak-like frequency dependence, as well as linear dichroism of inter Fermi arc transitions. Importantly, Fermi arc carriers have a strikingly different optical response as compared with DSM bulk carriers. The delineated optical response between bulk and Fermi arc are a Hallmark of how spectral weight is spatially distributed between bulk and surface in DSMs, and can be used to characterize DSMs optically.

\begin{acknowledgments}
We thank Alex Frenzel and Mark Rudner for useful conversations. This work was supported by the Singapore National Research Foundation (NRF) under NRF fellowship award NRF-NRFF2016-05.
\end{acknowledgments}

\onecolumngrid

\newpage

\section*{Supplementary Material for ``Large optical conductivity of Dirac semimetal Fermi arc surfaces states''}

\twocolumngrid

\subsection{Dispersion and Location of Fermi arc states}

The low energy effective Hamiltonian of 3D topological Dirac semimetal (DSM) in the bulk can be captured by a minimal $4\times4$ Hamiltonian $
\HH = \HH_0  + \HH' $
with 
\be
\HH_0 = m_{\vec q} \vec s_0 \boldsymbol{\tau}_3 + v (k_x \vec s_3 \boldsymbol{\tau}_1 - k_y \vec s_0 \boldsymbol{\tau}_2),
\quad
\mathcal{H}'\sim  \vec s_1,\vec s_2, 
\label{suppeq:fullhamiltonian}
\ee
where $\vec q = (k_x,k_y,k_z)$ is taken about the $\Gamma$ point, $\vec{s}_0$ ($\boldsymbol{\tau}_0$) is an identity matrix, and $\vec s_{1,2,3}$ ($\boldsymbol{\tau}_{1,2,3}$) are Pauli matrices representing the spin (orbital) degrees of freedom. We note that $m$ is a mass parameter that characterizes the band inversion of bands with different parity~\cite{wang,yang,neupanecd}. 
In three-dimensional topological Dirac semimetals, the crystal possesses rotational symmetry about the $\hat{\vec z}$ axis. Consequently, on the $k_z$ axis 
the band inversion takes the form
\be
m(k_x = 0, k_y = 0, k_z) = - m_0 + m_1 k_z^2,
\label{suppeq:mass}
\ee
and ${\cal H}'=0$. Using this form of the band inversion captured in $m(0,0,k_z)$ in Eq.~(\ref{suppeq:fullhamiltonian}), a pair of Dirac points $(0,0,\pm k_0)$ with $k_0 = \sqrt{ m_0 / m_1 }$ emerge in the bulk due to the band inversion around the $\Gamma$ point and rotational symmetry along the $k_z$ axis.

We will now consider a semi-infinite DSM that fills $y>0$, and vacuum for $y<0$. This leaves a $x$-$z$ surface at $y=0$. Note that in the DSM $m_0 > 0$ in Eq.~(\ref{suppeq:mass}) since bands with different parity are inverted close to $\Gamma$. As a result, we will model $m_0(y)>0$ for $y\geq0$ (for the DSM), and $m_0(y) < 0 $ when $y<0$ (for vacuum).

To obtain the Fermi arc surface states, we substitute $k_y \rightarrow - i \p_y$ and make an ansatz for the surface state wavefunction:
$ | \psi_{ \vec k \uparrow \downarrow} \ra = e^{ - y/\lambda_{ \vec k}} u_{\uparrow \downarrow} $ which is localized at the surface at $y=0$ and decays in the bulk. Here $\vec k = (k_x, k_y)$ and $u_{\uparrow \downarrow}$ are a pair of $k$-independent and orthogonal spinors corresponding to the $s=1$ and $s=-1$ spin blocks in Eq.~(\ref{suppeq:fullhamiltonian}). Taking $\HH' =0$ since it is weak and vanishes along the $k_z$ axis as $\mathcal{H}' \sim \mathcal{O}(k^3)$, we obtain an 
an eigenproblem
\be
{\cal H}_0 | \psi_{\vec k \uparrow \downarrow} \ra = \varepsilon_{\vec k \uparrow \downarrow} | \psi_{\vec k \uparrow \downarrow} \ra
\label{suppeq:eigenproblem}
\ee
for each of the spins. For spin $\uparrow$ ($\downarrow$), we have
\be
 \lp  \begin{array}{cc}   m_ {\vec k} - \varepsilon_{\vec k} & \pm v k_x + v / \lambda_{\vec k}  \\ \pm v k_x - v / \lambda_{\vec k} &   - m_ {\vec k} -\varepsilon_{\vec k}   \end{array}   \rp   \lp \begin{array}{c} \psi_{1(3)} \\ \psi_{2(4)} \end{array} \rp = 0,
\ee
where $\psi_{i}$ represents the components of the $4$-spinor $\psi = (\psi_{\vec k, \uparrow}, \psi_{\vec k, \downarrow})$; here $i=1,2,3,4$. 
Solving Eq.~(\ref{suppeq:eigenproblem}) yields Jackiw-Rebi type surface states localized around $y=0$ which decays into the bulk ($y\geq0$): 
\be
| \psi_{ \vec k \uparrow} \ra = \frac{e^{- y/\lambda_{\vec k}}}{\sqrt{2}} \left(\begin{array}{c} 1 \\ 1 \\ 0\\ 0\end{array}\right) , \quad | \psi_{ {\vec k} \downarrow} \ra = \frac{e^{-y/\lambda_{\vec k}}}{\sqrt{2}} \left(\begin{array}{c} 0 \\ 0 \\ 1\\ 1\end{array}\right)  ,
\ee
and energy
\be
\varepsilon_{ \vec k \uparrow \downarrow} = \pm v k_x , \quad  \lambda_{\vec k} = - v/ m_{\vec k} .
\label{suppeq:edgeconstraint} 
\ee
We emphasize that $\lambda_{\vec k}= - v / m_{\vec k}$ 
is a self-consistent equation since the mass parameter itself depends on $\lambda$ via $m(k_x, k_y = i / \lambda_{\vec k}, k_z)$. 

When $m_{\vec k} \rightarrow 0$, the decay length $\lambda_{\vec k} = - v / m_{\vec k}$ diverges and the wavefunction merges into the bulk. As a result, the positivity of the decay length, i.e., the normalizable requirement of the wavefunction, limits the existence of surface states in $k$-space to the region where $m_{\vec k} < 0$. Therefore we can only find Fermi arc states between $k_z \in [-k_0, k_0]$ with $k_0 = \sqrt{m_0/m_1}$, i.e., the two Dirac points.

\subsection{Effective Hamiltonian for the Surface States}

To obtain the effective Hamiltonian for Fermi arc surface states, we use $| \psi_{\vec k \uparrow \downarrow}  \ra$ as a basis and treat $\HH_{4 \times 4}'$ as perturbation. Projecting the full $4 \times 4$ Hamiltonian into this basis, we obtain the surface Hamiltonian
\be
\lb \HH_{\rm surf} \rb_{ij} =  \la \psi_{ \vec k i} |  {\cal H}_0 + {\cal H}'  | \psi_{ \vec k j} \ra , \quad (i= \uparrow \downarrow)   .
\ee

Using the basis wavefunctions from the previous section, one can verify that the effective surface Hamiltonian reads as
\be
\HH_{\rm surf,0} = v k_x \sigma_z + \Delta_{\vec k} \sigma_x   ,
\label{suppeq:surfhamiltonian}
\ee
where $\Delta_{\vec k}$ captures the coupling between the $\uparrow$ and $\downarrow$ surface states, and $\sigma_{x,y,z}$ are Pauli matrices.
For intrinsic inter-spin mixing captured by ${\cal H}'$, we find $\Delta_{\vec k} = \Delta_{\vec k}^{(0)} = \la \psi_{ \vec k \uparrow} | {\cal H}'  | \psi_{ \vec k \downarrow} \ra \sim  \la \psi_{ \vec k \uparrow} | k_z k_+^2 | \psi_{ \vec k \downarrow} \ra $, where $k_+ = k_x + i k_y$.

We note that $\Delta_{\vec k}^{(0)} $ has three nodes: one node at $\vec k =(0,0)$, and a pair at the projected Dirac points $\pm \vec k_0 = (0, \pm k_0)$. The former node, arises from a Kramers degeneracy of spin up and down surface states at $\Gamma$ point, and is protected by time-reversal symmetry. The later two nodes come from rotational symmetry along the $k_z$ axis. In the bulk and along the $k_z$ axis ($k_x=k_y=0$), $\HH_{4 \times 4}'= 0$ yielding no coupling between $\uparrow$ and $\downarrow$ states. Similarly, on the exposed surface $y=0$, we find $\Delta_{\vec k}^{(0)} = \alpha k_z (k_x - 1/\lambda_{\vec k} )^2 $. Crucially, 
at the two projected Dirac points $(k_x = 0, k_z =\pm \vec k_0)$, the inverse decay length $1 / \lambda_{\vec k_0} \rightarrow 0$. As a result, $\Delta_{\vec k}^{(0)} $ also vanishes at the pair of projected Dirac points due to rotational symmetry.

Surface potentials can also contribute to Eq.~(\ref{suppeq:surfhamiltonian}) yielding renormalized values of 
$\Delta_{\vec k} = \Delta_{\vec k}^{(0)} + \Delta'_{\vec k} $, where $\Delta'_{\vec k}$ 
arise from a surface potential $V_{\rm surf} $  mixes $\uparrow$ and $\downarrow$ spins on the Fermi arc states so that  
\be
\Delta'_{\vec k} = \la \psi_{\vec k \uparrow} | V_{\rm surf} | \psi_{\vec k \downarrow} \ra ,
\ee
where
\be
V_{\rm surf} = \left( \begin{array}{cc}  0 & V_{\uparrow \downarrow}  \\ V_{\uparrow \downarrow}^\dagger & 0  \end{array} \right)  ,  
\ee
with $V_{ss'}$ ($s,s'=\uparrow,\downarrow$) are $2 \times 2$ matrices. We note that for $V_{\rm surf} $ that breaks rotational symmetry, $\Delta_{\vec k} $ at $\pm \vec k_0 = (0, \pm k_0)$ may no longer vanish identically since this lifting the degeneracy of the two surface states at the projected Dirac nodes. However, for simplicity we will concentrate on $V_{\rm surf} $ that preserve rotational symmetry in the main text.

We emphasize that non-time-reversal-breaking surface potentials cannot lift the Kramers degeneracy at the $\Gamma$ point. To see this, note that
\begin{align}
\Delta(\Gamma) = & \la \psi_{\Gamma \uparrow} | {\cal H} (\Gamma) + V_{\rm surf} | \psi_{\Gamma \downarrow} \ra  \nn
= &  \la \psi_{\Gamma \uparrow} | {\cal H} (\Gamma) + V_{\rm surf} | \Theta \psi_{\Gamma \uparrow} \ra  \nn
= & \la (-i K s^y) \Theta \psi_{\Gamma \uparrow} | [{\cal H} (\Gamma) + V_{\rm surf}]^* | (-i K s^y)  \psi_{\Gamma \uparrow} \ra   \nn
= & \la \Theta^2 \psi_{\Gamma \uparrow} |  {\cal H} (\Gamma) + V_{\rm surf} | \Theta \psi_{\Gamma \uparrow} \ra    \nn
=& - \Delta(\Gamma) = 0  ,
\end{align}
during which we have used $[\Theta, {\cal H}(\Gamma) + V_{\rm surf} ] = 0$, and $\Theta = -i K s^y$ is the time-reversal operator where $s^y = (s^y)^\dagger$ acts on the real spin. From the second line to the the third line, we noted that $\Delta(\Gamma)$ is real valued and took the complex conjugate of overall expression. Lastly, we also noted that the time-reversal operator squares as $\Theta^2 = - 1$ (see fourth and fifth lines). To capture the both the intrinsic and surface potential contributions to inter arc mixing, we use a minimal phenomenological model for $\Delta_{\vec k}$ that respects both TRS and RS.

Lastly, we comment, parenthetically, that surface potentials can also in-principle induce particle-hole asymmetric terms. However, these can depend sensitively on the surface termination and sample preparation, and are beyond the current scope of our work.

\subsection{Jacobian factor for inter surface state transitions}

The Jacobian factor in the main text [see Eq.~(\ref{eq:particle-hole-spectrum})] are different for A-type and B-type contours. The idea is to avoid singularities by choosing different sets of parameters for A-type or B-type transitions.

In A-type contours, $\hbar \omega < 2 \Delta_0$, the contours are disconnected segments and can be described by the parameter $\phi_{\vec k}$ which is continuous between $-\pi$ to $\pi$. In this case, we have
\be
\sqrt{v^2 k_x^2 + \Delta_0^2 \sin^2 \frac{\pi k_z}{k_0} } = \eta, \quad \phi_{\vec k} = \arctan \frac{\Delta_0 \sin \frac{\pi k_z}{k_0}}{v k_x},
\ee
yielding 
\be
R = \left|\frac{\partial(k_x,k_z)}{\partial(\eta,\phi)}\right| = 2 \frac{1}{(2\pi)^{2}} \frac{k_0}{\pi v} \frac{ \eta}{ \Delta_0} \frac{1}{ [1 - (\eta / \Delta_0)^2 \sin^2 \phi ]^{1/2} },
\ee
 where a factor of $2$ in the numerator comes from adding up contributions from the central circle, as well as the side circles (see Fig.~\ref{fig1}b in the main text);
In B-type contours, $\hbar \omega \geq 2 \Delta_0$, the contours are two curves can be described by the parameter $k_z$ which is continuous between $-k_0$ to $k_0$. In this case, we use another set of parameters $(\eta, \nu)$ where $\nu \in [-\pi, \pi]$:
\be
\sqrt{v^2 k_x^2 + \Delta_0^2 \sin^2 \frac{\pi k_z}{k_0} } = \eta, \quad k_z = k_0 \frac{\nu}{\pi}  ,
\ee
and arrive at 
\be
R = \left|\frac{\partial(k_x,k_z)}{\partial(\eta,\nu)}\right| = 2 \frac{1}{(2\pi)^{2}} \frac{k_0}{\pi v}  \frac{1}{ [ 1 - ( \Delta_0 / \eta)^2 \sin^2 \nu ]^{1/2} }, 
\ee
in which the factor of $2$ results from summing over contributions from two singly-connected Fermi arcs (see Fig.~\ref{fig1}b in the main text).

\subsection{Optical conductivity of bulk states}

Neglecting cubic terms, the $4\times 4$ Dirac semimetal Hamiltonian~(Eq.~\ref{suppeq:fullhamiltonian}) has two non-interacting and time-reversal blocks, and each of them possesses a pair of Dirac nodes. We can pick one of the four Dirac nodes to calculate the optical response.  Near the Dirac nodes $(0,0, k_0)$, where $k_0 = \sqrt{m_0/m_1}$, the Hamiltonian can be simplified as
\be
\HH_{\uparrow} = (A_x k_x, - A_y k_y, A_z k_z)  ,
\label{suppeq:spinuphamiltonian}
\ee
where $A_x = A_y = v$, and $A_z = 2 \sqrt{m_0 m_1}$. The eigenstate and energy for the $2 \times 2$ Hamiltonian are
\be
\psi_{+  \vec q}(\vec r) = \lp \begin{array}{c} \cos \frac{\theta_{\vec q}}{2} \\ \sin \frac{\theta_{\vec q}}{2} e^{ i \phi_{\vec q}} \end{array} \rp ,
\quad
\psi_{- \vec q}(\vec r) = \lp \begin{array}{c} \sin \frac{\theta_{\vec q}}{2} \\ - \cos \frac{\theta_{\vec q}}{2} e^{i \phi_{\vec q}} \end{array} \rp ,
\label{suppeq:bulkwavefunction}
\ee
with
\be
\tan \theta_{\vec q} = \frac{A_x}{A_z} \frac{\sqrt{k_x^2 + k_y^2}}{k_z} , \quad
\tan \phi_{\vec q} = - \frac{k_y}{k_x} ,
\label{suppeq:thetaq}
\ee
and
\be
\xi^{\pm}_{\vec q } = \pm \sqrt{A_x^2 (k_x^2 + k_y^2) + A_z^2 k_z^2 }  ,
\label{suppeq:bulkdispersion}
\ee
where $\vec q = (k_x,k_y,k_z)$.

In the presence of incident light $\boldsymbol {\cal E} = \vec E \cos \omega t$, electron-hole transitions occur between the occupied bulk states and unoccupied bulk states. These can be described via the standard golden rule technique as
\be
W_\uparrow = \frac{2\pi}{\hbar} \sum_{\vec q} |M_{\uparrow,\vec q}|^2 \delta( \xi^{+}_{\vec q }- \xi^{-}_{\vec q } - \hbar \omega) \Theta(\hbar \omega - 2|\mu|) ,
\ee
where $\Theta(\hbar \omega - 2|\mu|)$ means that only when the energies of incident photons are greater than $|2\mu|$, which is the chemical potential counted from the Dirac nodes, the optical transitions are allowed.

Writing $\vec k \rightarrow \vec k - e \boldsymbol {\cal A}/ \hbar c$ in the spin up Hamiltonian (Eq.~\ref{suppeq:spinuphamiltonian}), with the vector potential satisfying $\boldsymbol {\cal A} = (1/c) \p_t \boldsymbol {\cal E}$, and expanding to linear order in $\vec E$ yields the matrix elements
\be
|M_{\uparrow, \vec q, i}|^2 = \lp \frac{i e A_i E_i }{2 \hbar \omega} \rp^2 | \la \psi_{+ \vec q} | \sigma^i | \psi_{- \vec q} \ra |^2,~(i=x,y,z) .
\ee 

Using the eigenstates given above, we have
\be
\la \psi_{+ \vec q} | \sigma^x | \psi_{- \vec q} \ra = \sin^2 \frac{\theta_{\vec q}}{2} e^{-i \phi_{\vec q}} - \cos^2 \frac{\theta_{\vec q}}{2} e^{ i \phi_{\vec q}} ,
\ee
\be
\la \psi_{+ \vec q} | \sigma^y | \psi_{- \vec q} \ra = i ( \sin^2 \frac{\theta_{\vec q}}{2} e^{-i \phi_{\vec q}} + \cos^2 \frac{\theta_{\vec q}}{2} e^{ i \phi_{\vec q}} ) ,
\ee
and
\be
\la \psi_{+ \vec q} | \sigma^z | \psi_{- \vec q} \ra = \sin\theta_{\vec q} .
\ee

For linearly polarized light $\boldsymbol {\cal E} = E \hat{\vec x} $ ($\boldsymbol {\cal E} = E \hat{\vec y} $), it gives
\be
| M_{\uparrow, \vec q,x(y)} |^2 = \lp \frac{e v E}{ 2 \hbar \omega} \rp^2 \lp \cos^2 \theta_{\vec q} \cos^2 \phi_{\vec q} + \sin^2 \phi_{\vec q}  \rp ,
\ee

The transition rates for both the Dirac nodes, $(0,0,\zeta k_0)$, are the same when they have the same chemical potential
\begin{align}
W_{\uparrow, x(y)} = & \frac{2\pi}{\hbar} \lp \frac{e v E}{ 2 \hbar \omega} \rp^2 \sum_{\vec q}  \lp \cos^2 \theta_{\vec q} \cos^2 \phi_{\vec q} + \sin^2 \phi_{\vec q}  \rp \nn
& \times \delta(\xi^{+}_{\vec q }- \xi^{-}_{\vec q } - \hbar \omega)~\Theta(\hbar \omega - 2|\mu|) \nn
= & \frac{e^2}{\hbar} \frac{1}{ 48 \pi} \frac{1}{A_z / L_z} L_x L_y  E^2~\Theta(\hbar \omega - 2|\mu|).
\end{align}

Using Ohm's law
\be
\hbar \omega W_{\uparrow, x(y)} =  \Re [ \sigma_{\uparrow, xx}^{\rm bulk}(\omega) ] E^2 L_x L_y / 2 ,
\ee
one has
\be
\Re [ \sigma_{\uparrow, xx}^{\rm bulk}(\omega) ] = \frac{e^2}{\hbar} \frac{1}{ 12 \pi} \frac{\hbar \omega}{A_z/L_z} \Theta(\hbar \omega - 2 |\mu |) ,
\ee
where we have summed over both two nodes.

Similarly one has
\be
\Re [ \sigma_{\uparrow, zz}^{\rm bulk}(\omega) ] = \frac{e^2}{\hbar} \frac{1}{ 12  \pi} \frac{\hbar \omega}{A_x^2 / A_z L_y}  \Theta(\hbar \omega - 2 |\mu |)  .
\ee

\subsection{Useful identity involving $\mathcal{I}_x$ and $I_x$}

During the calculation of interband conductivity described in the main text, the frequency dependence of interband conductivity arose from
\be I_x(\omega) = \frac{\bar{\omega}^{-\frac{3}{2} (1+\zeta)} }{4\pi} \int_{-\pi}^{ \pi } (1 - \bar{\omega}^{-2\zeta} \sin^2 \nu)^{-\frac{1}{2}} \sin^2 \nu   d \nu,
\label{suppeq:i}
\ee
where $\bar{\omega} = \omega / 2 \Delta_0$, $\zeta(\bar{\omega})= \sgn (\bar{\omega} -1)$.
Similarly, the intraband (Drude) conductivity exhibited chemical potential dependence in the form
\be
{\cal I}_x(\mu) = \frac{\bar{\mu}^{(1-\zeta)/2}}{4\pi} \int_{-\pi}^{\pi} \frac{ 1-\bar{\mu}^{-(1+\zeta)}\sin^2 \nu }{ (1-\bar{\mu}^{-2\zeta}\sin^2 \nu)^{1/2} } d \nu ,
\label{suppeq:cali}
\ee
where
$\bar{\mu} = \mu / \Delta_0$. In Eq.~(\ref{suppeq:cali}), the exponent $\zeta$ depends on the chemical potential as $\zeta(\bar{\mu})= \sgn (\bar{\mu} -1)$. We note that it will be useful to express both Eq.~(\ref{suppeq:i}) and ~(\ref{suppeq:cali}) in terms of the elliptic integrals of the first, and second kind, $K(x) = \int_0^{\pi/2} (1-x \sin^2 \phi)^{-1/2} d \phi$, and 
$E(x) = \int_0^{\pi/2} (1-x \sin^2 \phi)^{1/2} d \phi$ respectively [Note that here we define $K(x)$ and $E(x)$ in terms of the parameter $x = k^2$, where $k$ is the elliptic modulus]:
\be
I_x(\omega)  = \frac{ K(\bar{\omega}^{-2\zeta}) - E(\bar{\omega}^{-2\zeta}) }{\pi \bar{\omega}^{\frac{1}{2}(3-\zeta)} } ,
\label{suppeq:ellipticali}
\ee
and
\be
{\cal I}_x(\mu) = \frac{\bar{\mu}^{\zeta} E(\bar{\mu}^{-2\zeta}) - (-\bar{\mu} + \bar{\mu}^{\zeta}) K(\bar{\mu}^{-2\zeta})}{\pi \bar{\mu}^{\frac{1}{2}(1+\zeta)}}.
\ee
When $\zeta = \pm 1$, and by directly taking a derivative, we find
\be
\p_{\bar{\mu}} {\cal I}_x(\mu) = \frac{ K(\bar{\mu}^{-2\zeta}) - E(\bar{\mu}^{-2\zeta}) }{ \pi \bar{\mu}^{\frac{1}{2}(3-\zeta)} } , 
\label{suppeq:directderivative}
\ee
In obtaining Eq.~(\ref{suppeq:directderivative}) we have used the identities $
dK(x)/dx  = E(x)/2x(1-x) - K(x)/2x$, $dE(x)/dx = [E(x)  - K(x)]/2x$
Using $\p_{\mu} = \Delta_0^{-1} \p_{\bar{\mu}} $, and comparing Eq.~(\ref{suppeq:directderivative}) with Eq.~(\ref{suppeq:ellipticali}) we have
\be
\partial_{\mu} {\cal I}_x (\mu)= \Delta_0^{-1} I_x (2\mu),
\ee
where we have made the replacement of $\omega = 2\mu$ in the final line.


\begin{thebibliography}{99}
\bibitem{gauge} Here we adopted a continuous gauge that $\phi_{\vec k} \in [-\pi, \pi]$ with $\phi_{k_z = 0} = 0$.
\bibitem{dichroism} While ${\rm Re}(\sigma_{xx})$ exhibits a peak-like frequency dependence, ${\rm Re}(\sigma_{zz})$ features only a small kink at $\hbar \omega / 2 = \Delta_0$. This difference between ${\rm Re}(\sigma_{zz})$ and ${\rm Re}(\sigma_{xx})$ is due to the appearance of a factor of $(1- \bar{\omega}^{-2\zeta} \sin^2 \nu)$ in the matrix element $|{\cal M}_z|^2$ which eliminates the divergence in $\rho(\omega,\nu)\sim  (1- \bar{\omega}^{-2\zeta} \sin^2 \nu)^{-1/2}$ at $\bar{\omega} \rightarrow 1$ and $\nu \rightarrow \pi/2$.
\end{thebibliography}
\end{document}